\documentclass[twocolumn]{aastex631}

\newcommand\msun{{\rm M_{\odot}}}

\shorttitle{SSS Spin-down and Pulsation in Nova Her 2021}
\shortauthors{Drake et al.}

\begin{document}

\title{The Remarkable Spin-down and Ultra-fast Outflows of the Highly-Pulsed Supersoft Source of Nova Hercules 2021}

\correspondingauthor{Jeremy J.~Drake}
\email{jdrake@cfa.harvard.edu}

\author[0000-0002-0210-2276]{Jeremy J. Drake}
\affiliation{Harvard-Smithsonian Center for Astrophysics, 60 Garden St., Cambridge, MA 02138, USA}

\author[0000-0003-0440-7193]{Jan-Uwe Ness}
\affiliation{European Space Agency (ESA), European Space Astronomy Centre (ESAC), Camino Bajo del Castillo s/n, 28692 Villanueva de la Ca\~nada, Madrid, Spain}

\author[0000-0001-5624-2613]{Kim L. Page}
\affiliation{School of Physics \& Astronomy, University of Leicester, LE1 7RH, UK \\}

\author[0000-0002-2647-4373]{G. J. M. Luna}
\affil{CONICET-Universidad de Buenos Aires, Instituto de Astronom\'ia y F\'isica del Espacio (IAFE), Av. Inte. G\"uiraldes 2620, C1428ZAA, Buenos Aires, Argentina}
\affiliation{Universidad de Buenos Aires, Facultad de Ciencias Exactas y Naturales, Buenos Aires, Argentina.}
\affiliation{Universidad Nacional de Hurlingham, Av. Gdor. Vergara 2222, Villa Tesei, Buenos Aires, Argentina}

\author[0000-0001-5186-5950]{Andrew P. Beardmore}
\affiliation{School of Physics \& Astronomy, University of Leicester, LE1 7RH, UK \\}

\author[0000-0003-1563-9803]{Marina Orio}
\affiliation{Department of Astronomy, University of Wisconsin, 475 N. Charter Str., Madison WI 53706, USA}
\affiliation{INAF-Padova, vicolo Osservatorio, 5, 35122 Padova, Italy}

\author [0000-0002-1041-7542]{Julian P. Osborne}
\affiliation{School of Physics \& Astronomy, University of Leicester, LE1 7RH, UK \\}

\author[0000-0001-7016-1692]{Przemek Mr\'oz}
\affiliation{Division of Physics, Mathematics, and Astronomy, California Institute of Technology, Pasadena, CA 91125, USA}
\affiliation{Astronomical Observatory, University of Warsaw, Al. Ujazdowskie 4, 00-478 Warszawa, Poland}

\author[0000-0002-1359-6312]{Sumner Starrfield}
\affiliation{Earth and Space Exploration, Arizona State University, P.O. Box 871404, Tempe, Arizona, 85287-1404, USA
starrfield@asu.edu}

\author[0000-0003-4896-2543]{Dipankar P.~K. Banerjee} \affiliation{Physical Research Laboratory, Ahmedabad, INDIA 380009}

\author[0000-0001-6135-1144]{Solen Balman} \affiliation{Department of Astronomy and Space Sciences, Istanbul University, 34119 Istanbul, Turkey}
\affiliation{Faculty of Engineering and Natural Sciences, Kadir Has University, Cibali, 34083 Istanbul, Turkey}

\author[0000-0003-0156-3377]{M.~J.~Darnley} \affiliation{Astrophysics Research Institute, Liverpool John Moores University, IC2 Liverpool Science Park, Liverpool L3 5RF, UK}

\author[0000-0002-5967-8399]{\added{Y. Bhargava}}
\affiliation{Inter-University Centre for Astronomy and Astrophysics, Ganeshkhind, Pune, 411 007, India}

\author[0000-0003-1589-2075]{\added{G.~C. Dewangan}}
\affiliation{Inter-University Centre for Astronomy and Astrophysics, Ganeshkhind, Pune, 411 007, India}

\author[0000-0001-6952-3887]{\added{K.~P. Singh}}
\affiliation{Indian Institute of Science Education and Research Mohali, Sector 81, P.O. Manauli, SAS Nagar, 140 306, India}


\begin{abstract}
Nova Her 2021 (V1674 Her), which erupted on 2021 June 12, reached naked-eye brightness and has been detected from radio to $\gamma$-rays. An extremely fast optical decline of 2 magnitudes in 1.2 days and strong Ne lines imply a high-mass white dwarf. The optical pre-outburst detection of a 501.42~s oscillation suggests a magnetic white dwarf. This is the first time that an oscillation of this magnitude has been detected in a classical nova prior to outburst. We report X-ray outburst observations from {\it Swift} and {\it Chandra} which uniquely show: (1) a very strong modulation of super-soft X-rays at a \deleted{significantly} different period \replaced{to}{from} reported optical periods; \added{(2) strong pulse profile variations and the possible presence of period variations of the order of 0.1-0.3~s;} and (3) rich grating spectra that vary with modulation phase and show P Cygni-type emission lines with two dominant blue-shifted absorption components at $\sim 3000$ and 9000 km~s$^{-1}$ indicating expansion velocities up to 11000~km~s$^{-1}$. X-ray oscillations most likely arise from inhomogeneous photospheric emission related to the magnetic field. Period differences between reported pre- and post-outburst optical observations\added{, if not due to other period drift mechanisms,} suggest a large ejected mass for such a fast nova, in the range $2\times 10^{-5}$--$2\times 10^{-4} M_\odot$. A difference between the period found in the {\it Chandra} data and a reported  contemporaneous post-outburst optical period, \added{as well as the presence of period drifts,} could be due to weakly non-rigid photospheric rotation.
\end{abstract}


\section{Introduction} \label{s:intro}

Nova Herculis 2021 (TCP J18573095+1653396, ZTF19aasfsjq, hereafter V1674 Her) is proving to be one of the most fascinating nova events so far during this young century. It was discovered on 2021 June 12.5484 UT, at an apparent visual magnitude of 8.4 by Seidji Ueda (Kushiro, Hokkaido, Japan)\footnote{http://www.cbat.eps.harvard.edu/unconf/followups/\\J18573095+1653396.html}; an earlier detection of June 12.1903 UT was subsequently reported by the All-Sky Automated Survey for Supernovae (ASAS-SN; \citealt{Aydi.etal:21}). Later on the same day of discovery, the object was observed to reach a peak magnitude of approximately 6 \citep[][see also AAVSO Alert Notice\footnote{https://www.aavso.org/aavso-alert-notice-745}]{Munari.etal:21,Quimby.etal:21}, thus just visible to the unaided eye. 
Early spectroscopy indicated the object to be a very fast, reddened, classical nova \citep{Munari.etal:21,Aydi.etal:21}. 

V1674~Her subsequently garnered copious attention from radio to $\gamma$-rays.
\citet{Quimby.etal:21} 
found an extremely fast decay timescale, $t_2$, of 2 magnitudes in only 1.2~days using high-cadence optical photometry, and a remarkable 
plateau in the pre-maximum light curve at 8 magnitudes below peak that lasted for three or more hours. \citet{Li:21} found an uncatalogued gamma-ray source at the nova position based on Fermi-LAT data in the 0.1-300~GeV range obtained on 2021 June 12.0-13.3.
\citet{Sokolovsky.etal:21} reported the detection of V1674~Her on 2021 June 15-17 at radio frequencies.

Optical and infrared spectroscopy showed P Cygni profiles in addition to flat-topped Balmer line profiles with substantial substructure and an increasing line width with time, with widths up to 11000~km~s$^{-1}$ \citep{Albanese.etal:21,Balam.etal:21,Aydi.etal:21}. \citet{Woodward.etal:21} reported the appearance of infrared coronal emission lines based on data obtained on 2021 June 24, while \cite{Wagner.etal:21} observed strong \added{[Ne V] $\lambda 3426$ and [Ne III] $\lambda 3869,3968$} coronal lines on 2021 June 30 and concluded that V1674~Her is a neon nova originating on an ONe white dwarf (WD); based on its rapid photometric decline, \citet{Wagner.etal:21} also declared V1674~Her to be ``the fastest nova on record"\footnote{This indeed appears to be the case for Galactic novae, although we note that similarly fast candidates have been identified in M31 \citep[e.g.][]{Henze.etal:14}}.

A monitoring program by the {\it Neil Gehrels Swift Observatory (Swift)} began on 2021 June 13.5. The nova was detected by the X-ray Telescope (XRT)
on June 14.41 (day 2.22 after the ignition time of June 12.1903 based on the ASAS-SN data; \citealt{Aydi.etal:21}) and a supersoft X-ray source (SSS) was seen to emerge on July 1 \citep[day 18.9;][]{Page.etal:21}.
{\it Swift} UV observations were reported by \citet{Kuin.etal:21}.

Perhaps the most remarkable of the findings on V1674~Her is the detection of a 501.42~s periodic signal in archival $r$-band Zwicky Transient Facility (ZTF; \citealt{bellm2019}) data collected between 2018 March 26 and 2021 June 14 (i.e.~during pre-outburst quiescence) by \citet{Mroz.etal:21}. \citet{Shugarov.Afonina:21} and \citet{Patterson.etal:21} subsequently reported the detection of a 0.15302(2)d orbital period, with \citet{Patterson.etal:21} also finding a 501.52~s period. \citet{Mroz.etal:21} interpreted the 501s period as the spin period of a WD in an intermediate polar system (IP).

IPs host a magnetized WD with a magnetic field of 10$^{4-7}$~G and an accretion disk that is disrupted close to the WD by the magnetic field that channels accretion onto the polar caps from the disk at the magnetospheric radius
\citep[e.g.][]{Patterson:94,Wickramasinghe:14,Mukai:17}.
Thus V1674~Her is the first such object whose magnetic nature was possibly revealed early in the outburst. This discovery prompted a {\it Chandra} Director's Discretionary Time (DDT) High Resolution Camera (HRC-S) X-ray photometric observation and the subsequent detection by \citet{Maccarone.etal:21} of X-ray oscillations at a period of 503.9~s with an amplitude of 0.6 to 1.4 times the mean count rate. The oscillations were confirmed by  Neutron Star Interior Composition Explorer (NICER) observations that found a period of 
$501.8\pm 0.7$~s and that the count rate varied by up to a factor of $\sim 20$ over a 20 minute observation \citep{Pei.etal:21}.
This result places V1674~Her in a small group of \added{four} novae in which modulations of the SSS flux over the likely rotation period of the WD were observed \added{(V4743~Sgr, V2491~Cyg and V407~Lup being the other three. Periodic modulation has been seen in an additional handful of sources, but the nature of those variations remains less clear}. 

The distance to V1674~Her is presently not well-determined. \citet{Bailer-Jones.etal:21} list $d=6.0\pm ^{+3.8}_{-2.8}$~kpc based on {\it Gaia}-EDR3. A distance of $d\sim 4.75$~kpc was found by Woodward et al.\ (2021, submitted) using a variation of the maximum magnitude rate of decline (MMRD) method to derive the distance and extinction simultaneously. For the present, we adopt a nominal distance of 5~kpc.
 
Here, we present the first results of {\it Swift} X-ray monitoring observations and a second {\it Chandra}~DDT observation of V1674~Her, this time made at high spectral resolution using the Low-Energy Transmission Grating (LETG) and HRC-S detector. We report the {\it Swift} observation details in Sect.~\ref{s:swiftobs}, the {\it Chandra} observation and analysis in Sect.~\ref{s:obs}, before discussing and summarising the results in Sects.~\ref{s:discuss} and \ref{s:summary}.

\begin{figure*}
    \centering
    \includegraphics[angle=0, width=1\textwidth]{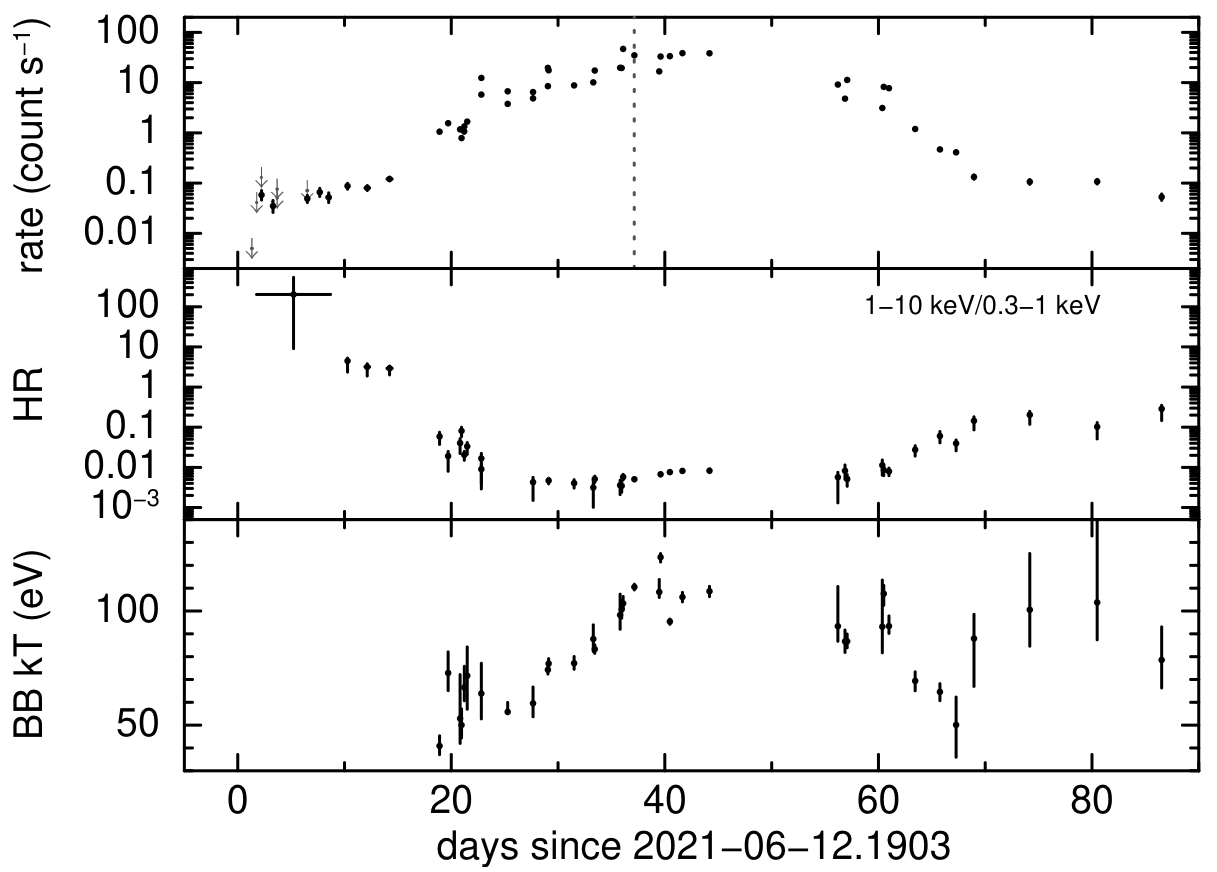}
    \caption{Swift-XRT 0.3--10~keV light-curve (top panel) and hardness ratio  comparing the 1--10 and 0.3--1~keV bands (middle panel); a hardness ratio value $<$~1 indicates that the SSS is dominating the emission. \added{The bottom panel shows the BB temperatures from the fit described in the text.} The arrows denote 3$\sigma$ upper limits. The vertical dotted line marks the time of the {\it Chandra} observation.}
    \label{f:xrtlc}
\end{figure*}

\section{{\it Swift} Observations and Analysis}
\label{s:swiftobs}

{\it Swift} observed V1674~Her typically every 1--2 days, from 1.3 days after the eruption, and continued up to day 44 (July 26); there was then a gap of 12 days, before regular observations resumed on day 56 (August 7). Given the optical brightness of the nova, most observations during the first ten days were taken using the Windowed Timing (WT) mode of the XRT to help minimise the effect of optical loading\footnote{https://www.swift.ac.uk/analysis/xrt/optical\_loading.php}. Photon Counting mode was then used from days 10--29, before the source count rate again increased and required WT mode. Only single pixel (grade 0) events were considered throughout. Data were analysed using HEASoft version 6.28 and the latest calibration files.

Spectra were extracted for each continuous snapshot of {\it Swift} data (typically 0.5--1~ks in duration) from the time when the SSS appeared (day 18.9) until the most recent observation, thus averaging over any variation caused by the periodicity. 

\subsection{X-ray Light Curve}
\label{s:swift_lc}

The early X-ray flux from V1674~Her was relatively faint and had a \added{(1-10 keV)/(0.3-1.0 keV)} hardness ratio HR~$>$~1, likely due to 
optically-thin emission. From day 18.9 a new supersoft component was seen below 1 keV. An exponential rise in the count rate followed, with superimposed short-term variations possibly due to the $\sim$~500~s periodicity (Fig.~\ref{f:xrtlc}). The soft X-ray flux peaked around day 40.

\subsection{Spectroscopic Analysis}
\label{s:swift_spectra}

Spectroscopic analysis of the {\it Swift} pulse-height spectra comprised model parameter estimation using the {\sc xspec} fitting engine. 
\deleted{In order to simplify the analysis, the spectra were fitted below 1~keV only, where the SSS emission strongly dominates. In this way, a simple model of an absorbed blackbody (BB) could be used to parameterize the emission, without the additional complication of including optically-thin emission components that would be necessary to model any (much fainter) underlying shock emission.} 
\deleted{A combination of a BB together with absorption edges at 0.55, 0.67, 0.74 and 0.87~keV (corresponding to N~VI, N~VII, O~VII and O~VIII, respectively) was used; this led to good fits, with reduced Cash statistic values of close to one. We also examined fits to the {\tt  rauch\_H-Ca\_solar\_90.fits} atmosphere grid\footnote{http://astro.uni-tuebingen.de/$\sim$rauch/TMAF/flux\_H-Ca.html}, which provided equally statistically acceptable results.}

\added{A combination of an optically-thin component (to model the underlying shock emission), and a BB together with absorption edges at 0.55, 0.67, 0.74 and 0.87~keV (corresponding to N~VI, N~VII, O~VII and O~VIII, respectively) was used to model the spectra during the SSS phase; the BB component strongly dominated during this time. This model led to good fits, with reduced Cash statistic values of $<$1.5.}

\added{We also examined fits to the {\tt  rauch\_H-Ca\_solar\_90.fits} atmosphere grid\footnote{http://astro.uni-tuebingen.de/$\sim$rauch/TMAF/flux\_H-Ca.html}, which provided equally statistically acceptable results.}

BB fits to the spectra indicated that the overall absorbing column remained constant, at a level of 2.9~$\times$~10$^{21}$~cm$^{-2}$, so fits were subsequently performed with N$_{\rm H}$ fixed at this value. Between days 18.9 and 27.7, the BB temperature remained approximately constant, at kT $\sim$54 ~eV; after this time, the emission steadily became hotter, reaching kT~$\sim$~130~eV ($1.5\times10^6$~K; the highest SSS temperature \added{for a Galactic nova} seen by {\it Swift} to date) on day \replaced{44.2}{39.6}. At some point during the observing gap between days 44 and 56, the SSS phase started to fade and cool, reaching a minimum of $\sim$~50~eV by day 67. After this time, the temperature started to increase again, returning to $\sim$~100~eV by day 74.2. \added{The BB temperatures are shown in the bottom panel of Fig.~\ref{f:xrtlc}}.

During the SSS phase, the dominant absorption edges required to improve the BB fit changed. At the beginning, when the SSS temperature was lower, the edges at 0.55 and 0.67~keV were a significant improvement on the BB-only fit. As the BB temperature started to increase, these became less important, with the higher energy 0.74 and 0.87~keV edges starting to dominate. We note that the trend of increasing and then decreasing BB temperature was also found if the spectra were fitted without absorption edges, and when using atmosphere model fits.

Comparison between the {\it Chandra} grating data (Sect.~\ref{s:obs}) and the approximately contemporaneous XRT observation shows a very similar overall spectral shape.

\section{{\it Chandra} Observations and Analysis}
\label{s:obs}

V1674~Her was observed by {\it Chandra}  using the LETG+HRC-S grating and detector combination. The observation began at 2021 July 19 UT07:40 (day 37.3 of the outburst)
for a net exposure time of 29641~s. Data were analysed using CIAO software version 4.13 with calibration database CALDB version 4.9.5\footnote{https://cxc.harvard.edu/ciao/}. 

\subsection{Timing Analysis}
\label{s:timing}

\begin{figure*}
    \centering
    \includegraphics[width=0.97\textwidth]{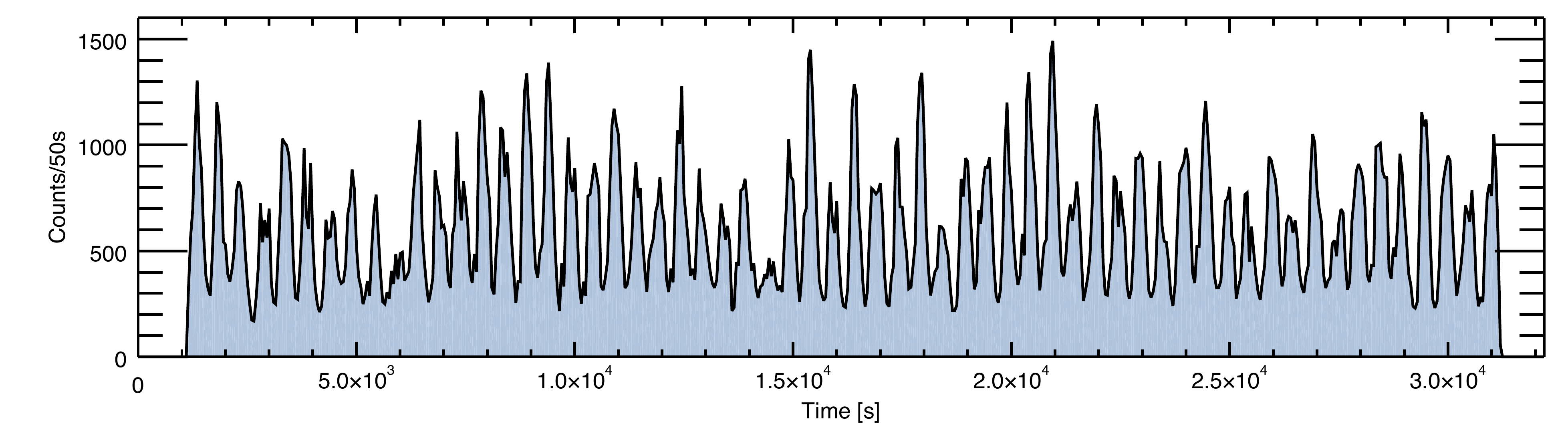}
    \includegraphics[width=1\textwidth]{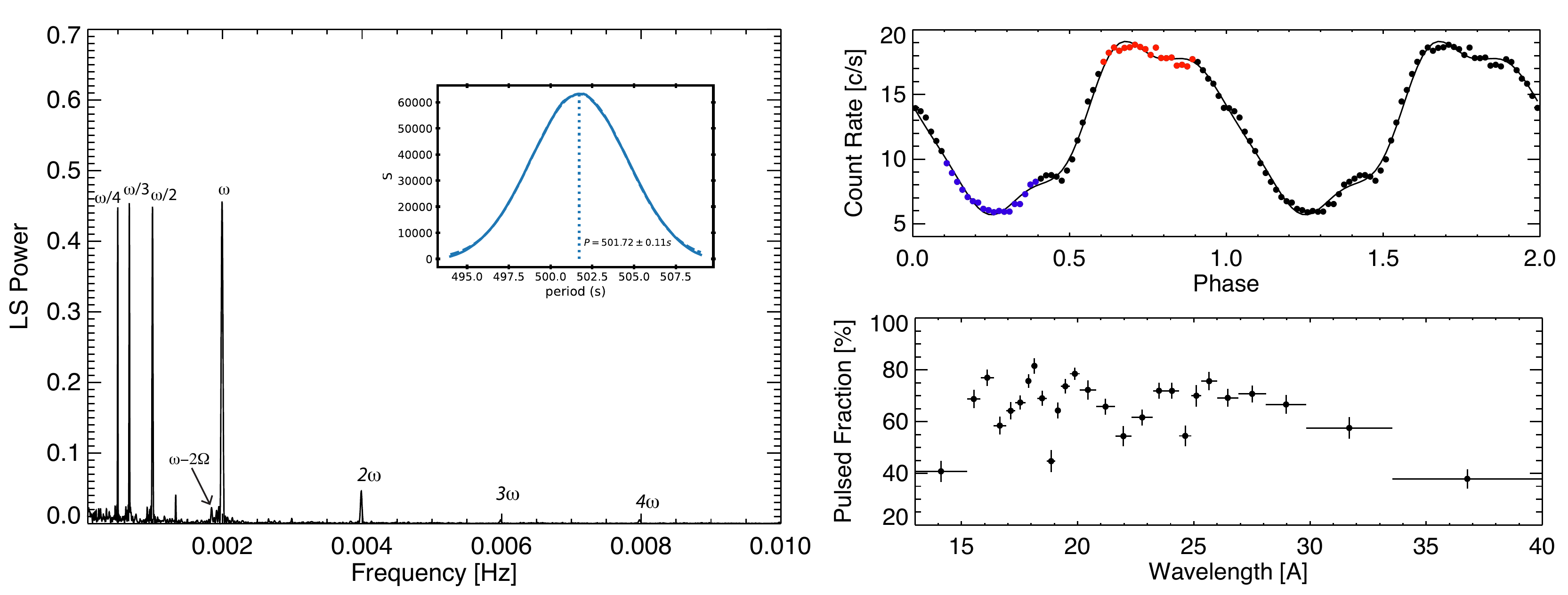}   
    \caption{The timing analysis of the LETG+HRC-S observation of V1674~Her. {\it Top}: The HRC-S 0th-order light curve in 50~s bins. {\it Bottom Left}: The LS power spectrum of V1674~Her derived from the HRC-S 0th-order light curve sampled with 1~s binning with some significant frequencies highlighted, including the $\omega$-2$\Omega$ beat period reported by Patterson et al. (2021). The inset shows the $\chi^2$ vs Period plane centered on the strongest peak of the power spectrum and a Gaussian fit to it. {\it Bottom  Right/Top}: The light curve folded at a period of 501.72 s, with 40 bins/cycle and a fit of a model comprising a series of sine waves whose frequencies are integer multiples of the main frequency $\omega$. Note that the error bars are smaller than the symbol size. The phase ranges for states deemed ``high'' and ``low'' for which phase-dependent spectra are extracted in Sect.~\ref{s:spectra} are highlighted with red and blue symbols, respectively.  {\it Bottom  Right/Bottom}: The pulsed signal fraction (defined as $(max-min)/max$ in the count rate light curve) in different wavelengths derived from the dispersed $\pm1$~order signal.
    }
    \label{f:lc}
\end{figure*}

Light curves were first constructed from both the 0th order and the $\pm$1 order ($\texttt{tg\_m=1,-1}$) signals after applying barycentric corrections to photon arrival times. The extraction region for 0th order was a circle of radius 3.8~arcsec while the default source and background region were used in the case of the $\pm 1$ orders. Background was negligible in comparison to the source signal. The 0th order lightcurve is illustrated in Figure~\ref{f:lc}.

In order to search for periodic signals, including near the $\sim 501.42$~s signal found in the ZTF data \citep{Mroz.etal:21}, we employed 
a Lomb-Scargle periodogram \citep[LS,][]{Lomb:76,Scargle:82}
and epoch-folding.  To determine the uncertainty in the period, 1000 light curves comprising a constant and sine wave with the same pulsed fraction, count rate, and 30~ks duration as observed by {\it Chandra} were analysed. Each simulation corresponded to a different Poisson realisation of the model. The recovered period had a standard deviation of 0.11s.

Both 0th and 1st order events were analysed separately, each yielding the same result. The best-fit period from the 0th order data was $501.72\pm 0.11$~s and is shown in Fig.~\ref{f:lc}. This period is also close to the 503.9~s period reported for the initial {\it Chandra} HRC-S observation obtained on 2021 July 10 UT01:12 by \citet{Maccarone.etal:21} and the 501.8$\pm$0.7 s period reported from a NICER observation obtained between 2021 Jul 10 and 2021 July 12. \citep{Pei.etal:21}. However, as we discuss later in Sect.~\ref{s:disc_period}, it is also \replaced{significantly}{likely} different from the ZTF period reported by \citet{Mroz.etal:21}, and the period found by \citet{Patterson.etal:21}

The 0th order light curve folded at the 501.72~s period is illustrated in Fig.~\ref{f:lc} and shows significant departures from a pure sine wave. A 4th-terms LS periodogram shows power at $\omega$, $2\omega$, $3\omega$ and $4\omega$ with aliases of the main period at $\omega/2$, $\omega/3$ and $\omega/4$.  
We fit the light curve using a model including these harmonics in a Fourier series and sinusoidal components with frequencies that are integer multiples of the main frequency $\omega$ \citep{VanderPlas2018}.   

\added{The phase-folded light curve through the {\it Chandra} observation shows large variations in the pulse profile. This is also evident in the light curve in Figure~\ref{f:lc}. Using similar methods to the period search described above, we also searched for period variations within the observation to probe for period drifts or quasi-periodic oscillations (QPOs). Suggestive period variations were found at the level of $\sim 0.1$-0.3s. However, we were unable to verify their significance as variations were of a similar magnitude to uncertainties.}

Phased light curves were also extracted for different wavelength ranges from the dispersed spectral orders. The pulsed fraction as a function of wavelength is also illustrated in Fig.~\ref{f:lc}.


\begin{figure*}
    \centering
    \includegraphics[width=0.95\textwidth]{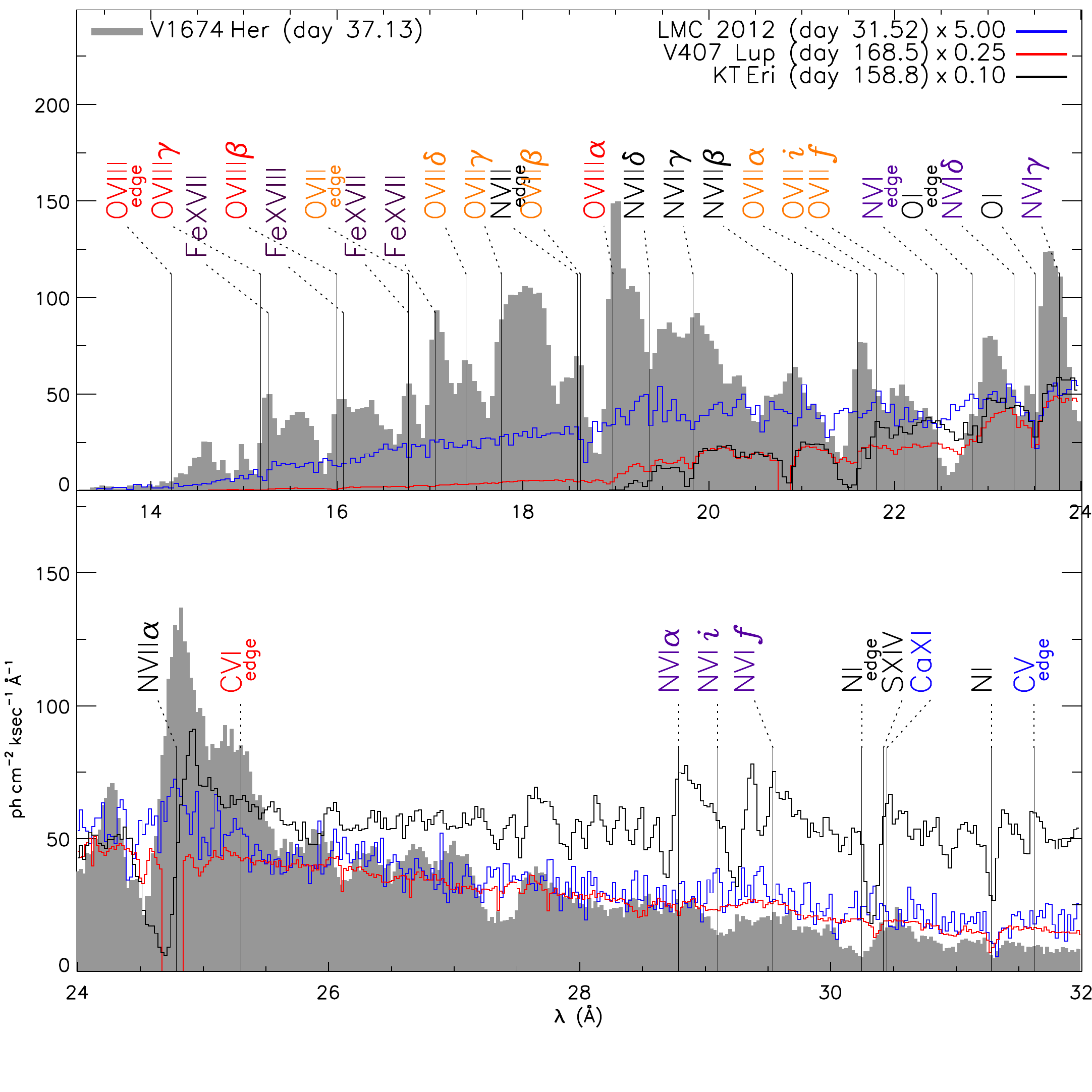}
    \includegraphics[width=0.95\textwidth]{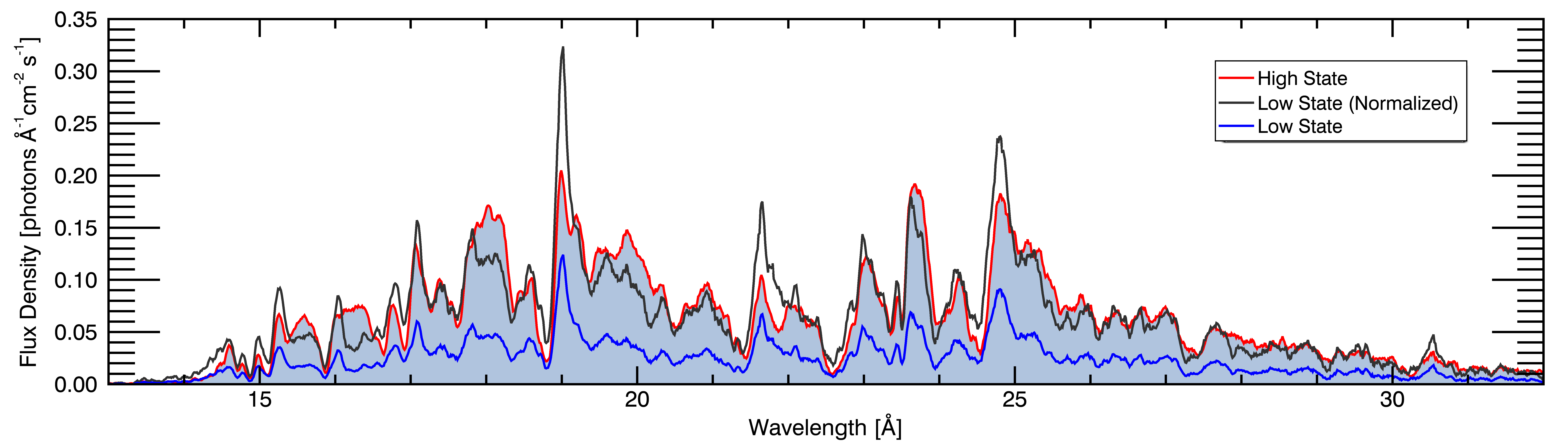}
    \caption{The {\it Chandra} LETG+HRC-S spectrum of V1674~Her. {\it Top}: Rest wavelengths of prominent spectral lines are noted in different colors for the different ions, and, to put V1674~Her (grey shadings) into context, overplotted are other bright nova SSS spectra.  
    {\it Bottom}:  The spectra extracted for ``high'' and ``low'' phases indicated in the phased lightcurve shown in Figure~\ref{f:lc}. Also shown is the ``low" state spectrum renormalized to the same total flux as the ``high'' state spectrum.
    }
    \label{f:cmp_v1674her}
\end{figure*}



\subsection{Spectroscopic Analysis}
\label{s:spectra}

The first order spectrum was extracted from the event list using the CIAO {\tt tgextract} routine. Owing to the fairly narrow wavelength range of the bright spectrum, higher-order contamination of the first order spectrum was negligible. 

In the top panel of Fig.~\ref{f:cmp_v1674her}, the spectrum of V1674\,Her is shown with gray shadings. To put this spectrum into context, additional Chandra LETGS spectra of other bright nova SSS are overplotted for comparison: LMC\,2012 (ObsID 14426), V407 Lup (ObsID 20632), and KT Eri (ObsID 12203), scaled with the factors given in the top right legend. The positions of some prominent spectral lines and ionization energies (edges) have been included.
In the bottom panel, spectra extracted for the ``high'' and ``low'' phases highlighted in the phased light curve in Figure~\ref{f:lc} are shown.

\section{Discussion}
\label{s:discuss}

\subsection{V1674 Her is a very fast nova}

With an optical decline from peak by 2 magnitudes in 1.2 days \citep{Quimby.etal:21}, V1674 Her is classed as a very fast nova (one of the fastest known). 
Such rapid fading is commonly ascribed to a comparatively low mass of ejected gas ($10^{-6}$-$10^{-5} M_\odot$) at high velocity (up to 10000~km~s$^{-1}$; \citealt[e.g.][]{Gehrz.etal:98}), which itself arises from a low accreted mass at the time of eruption; 
for a low accreted mass, a high mass WD is needed to achieve the pressure necessary for nuclear ignition \citep[][and references therein; see also \citealt{starrfield_2020_aa}]{Starrfield.etal:16}. A high WD mass is also indicated by the detection of optical Ne lines, leading \citet{Wagner.etal:21} to propose the V1674~Her is an ONe WD, which is expected to have a mass $>$ 1.05 $\msun$ \citep[][but see also \citealt{Althaus.etal:21}]{doherty_2015_aa}. 

A low ejecta mass and high ejecta velocity can be expected to lead to a rapid fall in the column density to the WD photosphere and thus a rapid emergence of the SSS, as was observed. First seen at 18.9 days after outburst, the SSS is in the fastest quartile to emerge among the Swift novae \citep{schwarz_2011_aa, Page.etal:20}.  

The normalizations to model atmosphere and BB fits to the {\it Swift} spectra around the time of the {\it Chandra} observation indicated luminosities of approximately $L\sim 1 \times 10^{38}$~erg~s$^{-1}$ and $ 3\times 10^{37}$~erg~s$^{-1}$, respectively, for a 5~kpc distance. A luminosity derived using a BB approximation to the SSS spectrum must be considered uncertain. However, the consistency between the BB and atmosphere model values to within a factor of a few lends modest support to the order of magnitude derived. The SSS emission is then essentially at the Eddington Limit ($L=1.26\times 10^{38}$~erg~s$^{-1}$ for a $\sim 1M_\odot$ WD; the actual limit will be lower for SSS emission because the gas is not fully-ionized).

\subsection{Phase-dependent X-ray emission}
\label{s:disc_phase}

There are two potential explanations for the periodic modulation of the SSS X-ray emission. As noted in Sect.~\ref{s:intro}, \citet{Maccarone.etal:21} have interpreted the modulation as being due to the spin period of the WD. Before addressing that, we also note that 
g-mode pulsations could be responsible. Indeed, some hints of g-mode pulsations during the SSS phase of other novae were reported by \citet{Drake.etal:03} and \citet{Leibowitz.etal:06}, and the periodicity that is present both before and after the TNR, is a reasonable g-mode period \citep{Starrfield.etal:84}.

In the case of novae, partial ionization zones of carbon and oxygen near the surface might potentially drive pulsations; this is the mechanism in the PG~1159 stars \citep{Starrfield.etal:84}. 
However, the temperature of the V1674~Her WD is far hotter than the pulsating WDs and the amplitude of the oscillations is much greater than in well-known pulsating WDs \citep{Corsico.etal:19}. Moreover, while a small change in pulsation period pre- and post-outburst appears to have occurred (Sect.~\ref{s:disc_period} below), the enormous change in photospheric temperature and structure with such a small period change makes a g-mode pulsation explanation unlikely. We conclude that WD spin modulation, originating due to the strong magnetic field, remains the most likely explanation for the pulsed emission.


V1674~Her is then \added{probably} a member of a very small group of novae that have shown modulations in the SSS flux at the spin period of the WD.
Three other novae, namely V4743~Sgr \citep{Ness.etal:03, Leibowitz.etal:06, Zemko.etal:17, Dobrotka.etal:17}, V2491 Cyg  \citep{Ness.etal:11, Zemko.etal:15} and V407 Lup \citep[][and Orio et al. in preparation]{Aydi.etal:18}, have shown pulsations with timescales of the order of tens of minutes during the SSS phase, and have later exhibited the same modulations in X-ray flux due to accretion at quiescence. This is typical of IPs. The transient M31 SSS described by \citet{King.etal:02} likely belongs to the same group. 

Modulation in the SSS at the WD spin period indicates the presence of either inhomogeneous surface emission, or inhomogeneous absorption. The latter might be expected from material co-rotating and constrained by the strong magnetic field, such as an accretion stream. A pure absorption source of modulation seems unlikely based on the wavelength-dependent pulsed fraction shown in Fig.~\ref{f:lc}. The pulsed fraction is smaller toward longer wavelengths, which is opposite the trend expected for absorption unless very high ionization renders the longer wavelengths transparent.

The modulation then seems most likely due to an inhomogeneous photosphere. Such a variation in emission might potentially be caused by ongoing accretion causing temperature variations over the photospheric surface (possibly due to the burning of accreted fresh H-rich fuel), or the strong surface magnetic field affecting the photosphere and perhaps the nuclear burning shell beneath. Accretion heating is ruled out because the pulsed fraction is large and the SSS appears close to the Eddington luminosity, which is orders of magnitude higher than the accretion luminosity prior to outburst.

\subsection{High-resolution SSS Spectrum}

Comparisons with previous LETG+HRC-S spectra of particularly bright SSS are shown in Fig.~\ref{f:cmp_v1674her}. 
The ``Wien tail'' high-energy cut-off in principle provides a first-order estimate of the effective temperature and indicates V1674~Her is hotter than the other novae shown,
although complications due to the 
O\,{\sc viii} absorption edge in V1674~Her render the comparison more uncertain.


In the soft range, the effects of $N_{\rm H}$ dominate. We note in passing that KT\,Eri had lower $N_{\rm H}$, even lower than LMC 2012, while V1674~Her has a similar flux decline with wavelength to other SSS spectra with $N_{\rm H}\sim 10^{21}$\,cm$^{-2}$. 

V1674~Her has  conspicuous P Cygni-like line profiles while in RS~Oph, for example, absorption lines were deeper and less blue-shifted. The most prominent P~Cyg profile is the O\,{\sc viii} Ly$\alpha$ transition near 19~\AA, with the Ly$\beta$ and possibly Ly$\gamma$ transitions also in evidence. The O\,{\sc viii} absorption lines are clearly blue shifted with respect to the labels (placed at rest-wavelength), more so than in V407~Lup and KT~Eri but less than in LMC~2012. The He-like O\,{\sc vii} line is also detected at 21.6\,\AA\ with a P Cyg profile 
The intercombination and forbidden lines at 21.8 and 22.1\,\AA\ are not seen in emission suggesting the resonance line at 21.6\,\AA\ is  photoexcited.
The N\,{\sc vi} He-like lines are not clearly detected.

The spectra show significant variations with phase. The pulsed fraction vs.\ wavelength (Fig.~\ref{f:lc}) is smallest in the cores of strong absorption lines, indicating saturation.
The lower panel in Fig.~\ref{f:cmp_v1674her} reveals comparatively much stronger emission line components in the ``low'' state spectrum relative to the ``high'' state. Absorption features are largely similar between states. 
The low state has comparatively more flux shortward of about 15~\AA, indicating that the temperature could be somewhat higher than in the high state. 

\begin{figure}
    \centering
    \includegraphics[width=0.47\textwidth]{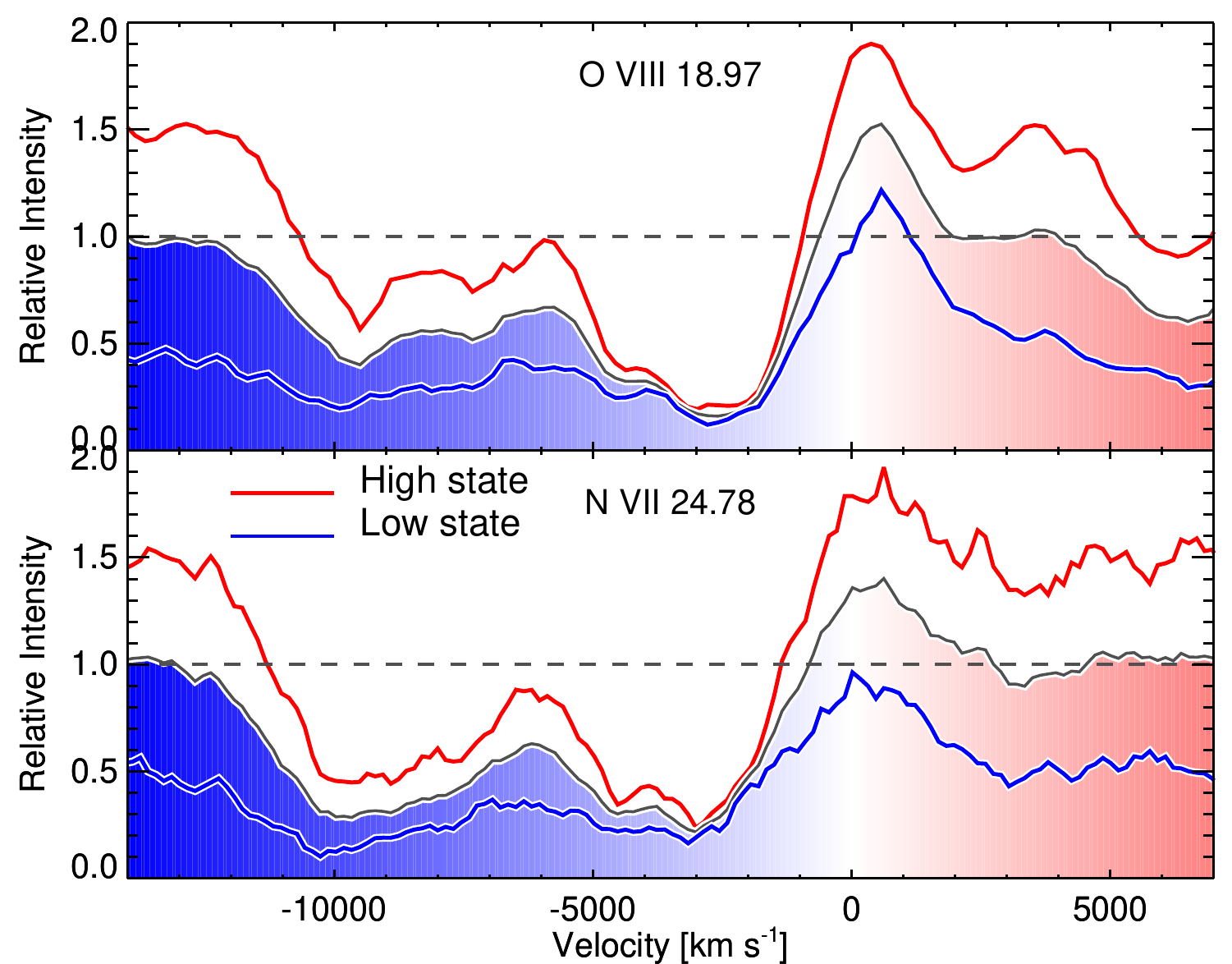}
    \caption{The Ly$\alpha$ lines of hydrogenic O and N demonstrating P Cygni behaviour, with structured blueshifted absorption out to velocities of 11000~km~s$^{-1}$ or so. The spectrum from the entire observation is shown (blue-red background) in addition to spectra corresponding to the ``high'' and ``low'' states.}
    \label{f:linevel}
\end{figure}

Classic P Cyg profiles comprise an  emission line component formed in a dense, radiatively-driven outflow close to the stellar photosphere, while the blueshifted absorption arises because the radiation passes through the outflow material rapidly expanding in the direction of the observer. 
Fig.~\ref{f:linevel} shows the O\,{\sc viii} and N\,{\sc vii} Ly$\alpha$ transitions in velocity space. The emission peaks are noticeably redshifted, confirming an origin behind or within the fast outflow.

The blueshifted absorption in Fig.~\ref{f:linevel} has considerable structure with two dominant components visible in both high and low states centered on velocities of approximately 3000 and 9000~km~s$^{-1}$. These profiles 
demonstrate that the outflow is inhomogeneous and expanding at remarkable speeds up to 11000~km~$^{-1}$, similar to the velocity broadening seen in Balmer line profiles \citep[e.g.,][]{Aydi.etal:21}.

\subsection{Detection of a Spin Period Change of the White Dwarf?}
\label{s:disc_period}

The ZTF data (obtained pre-outburst 2018 March 26 to 2021 June 14) indicate an optical rotation period of $P=501.4277\pm 0.0002$~s \citep[][and reanalysis of those data for this work]{Mroz.etal:21}. This is \deleted{significantly} different from the X-ray period derived here, $P=501.72\pm 0.11$~s. Similarly, the optical period of \citet{Patterson.etal:21} (post-outburst from data between 2021 July 1--August 10) of $501.52\pm 0.02$~s, is significantly longer than the ZTF period and also \deleted{significantly} shorter than the {\it Chandra} period.

It is tempting to interpret the periods in terms of rigid rotation and angular momentum loss resulting from the outburst. However, the \citet{Patterson.etal:21} data straddle the {\it Chandra} observation, indicating \replaced{X-ray and optical periods are different.}{that either X-ray and optical periods are different, or that period variations hinted at by the analysis in Sect.~\ref{s:timing} are real. We investigate these possibilities below.} 

\subsubsection{Period change under rigid rotation}

The \citet{Mroz.etal:21} and \citet{Patterson.etal:21} periods \added{at face value} imply a change of $\Delta P=0.09\pm 0.02$~s or $\Delta P/P\sim 10^{-4}$, which might be attributed to the outburst.
\added{If period drift or a QPO is not to blame,} there are two mechanisms through which this might have happened: an increase in scale height of a substantial portion of the WD envelope related to the TNR and a consequent increase in moment of inertia; or the loss of mass in the explosion that carried away angular momentum. 


The photospheric radius of a novae SSS is known to be expanded relative to that of the underlying WD by factors as much as 10 \citep[e.g.][]{Balman.Gamsizkan:17}.  
While the structure and mass of the expanded envelope are highly uncertain, by approximating the envelope as a spherical shell and dropping factors of approximately unity we can estimate the fractional moment of inertia change as
\begin{equation}
    \frac{\Delta I}{I}\approx M_{env}(\varepsilon^2 -1),
    \label{e:moi}
\end{equation}
where $M_{env}$ is the mass of the expanded envelope in solar masses, $\varepsilon$ is the factor of expansion of the bulk of the mass relative to the quiescent white dwarf radius, and for simplicity we have assumed a WD mass $M_{WD}\approx M_\odot$ ($M_{WD}$ could be as high as $1.35 M_\odot$). The envelope mass could be in the range $10^{-6}$--$10^{-4} M_\odot$, depending on the accreted mass required for TNR and the amount of mixing with underlying material \citep[e.g.,][]{Gehrz.etal:98}. Taking a large expansion factor, e.g., $\varepsilon=5$, and assuming rigid rotation (see Sect.~\ref{s:lima} below), fractional period changes of $10^{-4}$ could then potentially be explained by a simple moment of inertia change.

The angular momentum loss rate, $dJ/dt$, resulting from mass loss can be written \citep[e.g.,][]{Kawaler:88}
\begin{equation}
    \frac{dJ}{dt}=\frac{2}{3}\frac{dM_{ej}}{dt}R_{WD}^2\Omega\left(\frac{r_A}{R_{WD}}\right)^n,
    \label{e:kawaler}
\end{equation}
where $dM_{ej}/dt$ is the ejected mass loss rate, $R_{WD}$ is the WD radius, $n$ is related to the complexity of the magnetic field ($n=2$ for a radial field and $3/7$ for a dipole in the \citealt{Kawaler:88} formalism) and $r_A$ is the Alfv\'en radius, at which point the outflow exceeds the local Alfv\'en speed and is no longer forced to co-rotate by the magnetic field.  

Assuming the ejected mass is a smaller fraction of the total WD mass than the fractional change in rotation period, we can ignore the change in moment of inertia of the WD due to mass loss. 
Further assuming for simplicity that the mass loss rate is constant ($R_A$ otherwise depends on $\dot{M}_{ej}$), from Eqn.~\ref{e:kawaler} conservation of angular momentum then gives 
\begin{equation}
    \frac{2}{3} M_{ej} R_{WD}^2 \left(\frac{r_A}{R_{WD}}\right)^n =
    I_{WD}\frac{\Delta\Omega}{\Omega},
\end{equation}
where $\Delta\Omega$ is the change in angular velocity. 
For a dipolar field ($n=3/7$) and the observed period change, and writing $I_{WD}=\alpha M_{WD}R_{WD}^2$ and eliminating the factor of order unity, we have for the required Alfv\'en radius
\begin{equation}
 \frac{r_{A}}{R_{WD}}\sim \left( 10^{-4} \alpha \frac{M_{WD}}{M_{ej}}\right)^{7/3}.
 \label{e:alfrad_mej}
\end{equation}
For a uniform sphere, $\alpha=0.4$, but is considerably lower for a centrally-condensed WD and depends on mass. Adopting $\alpha\sim 0.2$ \citep[e.g.,][]{Roy.etal:21,Boshkayev.etal:17}, and for $M_{WD}\approx 1M_\odot$ the Alfv\'en radius is $r_A/R_{WD}\sim 2\times 10^{-5}/M_{ej} $. Thus for $r_A/R_{WD}$ of order unity, the implied ejected mass is of the order of $10^{-5}M_\odot$, with the required $r_A/R_{WD}$ growing rapidly for smaller ejected mass. 


\begin{figure}
    \centering
    \includegraphics[width=0.47\textwidth]{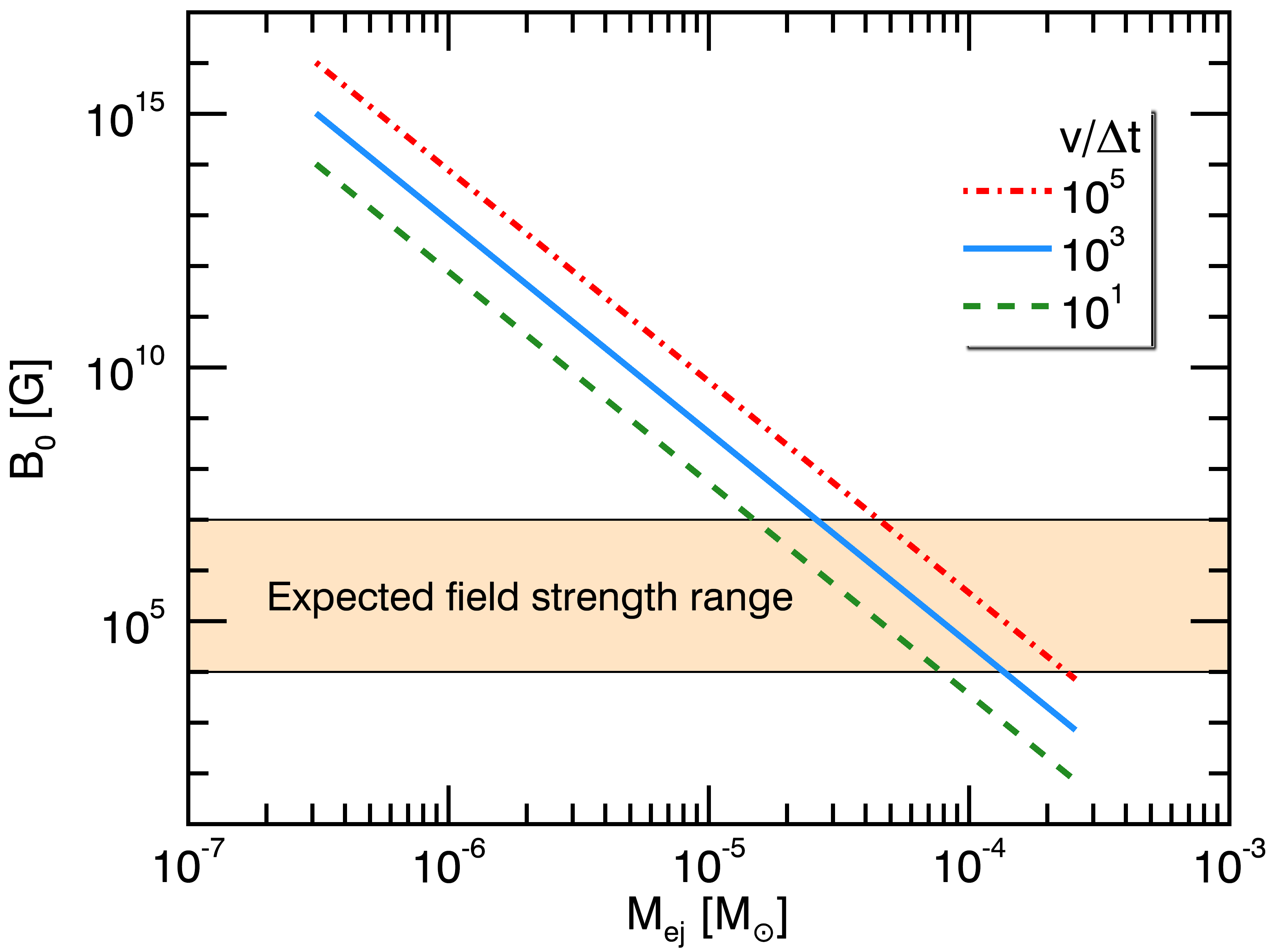}
    \caption{The WD surface magnetic field strength as a function of the ejected mass for different values of the ratio of outflow velocity at the Alfv\'en radius and the interval over which mass loss occurred, $v/\Delta t$. }
    \label{f:b0_mej}
\end{figure}

The Alfv\'en radius can also be written in terms of the surface magnetic field strength, $B_0$, and the mass loss rate, $\dot{M}$, and outflow velocity, $v$, as (e.g., adopting Eqn.~4 of \citealt{Kawaler:88} to the dipole case)
\begin{equation}
   \frac{r_A}{R_{WD}} =(B_0R_{WD})^{1/2} \left(\frac{1}{\dot{M}v}\right)^{1/4}.
   \label{e:alfrad_b0}
\end{equation}
Combining Eqns.~\ref{e:alfrad_mej} and \ref{e:alfrad_b0} and writing $\dot{M}=M_{ej}/\Delta t$, with $\Delta t$ being the interval over which $M_{ej}$ is lost, we can determine the required surface field strength, $B_0$, as a function of $M_{ej}$,
\begin{equation}
    B_0=\frac{1}{R}(10^{-4}\alpha M_{WD})^{14/3}\left(\frac{v}{\Delta t}\right)^{1/2}\frac{1}{M_{ej}^{25/6}}.
\end{equation}
This is illustrated in Figure~\ref{f:b0_mej} for different values of $v/\Delta t$; as a reference, for an outflow at 1000~km~s$^{-1}$ and an ejection timescale of the order of a day ($\sim 10^5$~s), $v/\Delta t\sim 10^3$.  

We noted above that, in the case of a fast nova, we might expect a low ejected mass (say $M_{ej}\sim 10^{-6}M_\odot$; e.g., \citealt[][]{Gehrz.etal:98}). In contrast,  Figure~\ref{f:b0_mej} demonstrates that for typical IP magnetic field strengths an ejected mass in the range $2\times 10^{-5}$--$2\times 10^{-4} M_\odot$  is required. Such a mass is not unprecedented: \citet{Vanlandingham.etal:96} estimated $M_{ej}\sim 1.8 \times 10^{-4} M_\odot$ in the case of the fast nova V838~Her ($t_2 \sim 2d$).

\subsubsection{Period change through a LIMA analogy}
\label{s:lima}

The difference in \citet{Patterson.etal:21} and {\it Chandra} periods, \added{and the possible presence of a QPO as noted in Sect~\ref{s:timing},} lead us to consider a third period change mechanism: that the effective photosphere post-outburst is not rigidly rotating with the underlying WD, but is at or slightly beyond $r_A$ where the magnetic field can no longer enforce strict co-rotation. The optical photosphere lying closer to the WD can then have a shorter rotation period than the X-ray photosphere\added{, and both periods could be subject to drift}. The observed rotation period could \replaced{then}{also} change through the outburst and SSS phase according to the mass loss rate and changing photospheric radius, with expectation of a longer period earlier in the SSS phase when mass loss rates in the radiatively-driven outflow would be larger.

Such a mechanism is reminiscent of the Low-Inertia Magnetic Accretor (LIMA) model \citep{Warner.Woudt:02} in which accretion onto an equatorial belt of the WD causes the belt to vary its angular velocity when the magnetic field is insufficiently strong to enforce rigid rotation.

\section{Summary}
\label{s:summary}

High-resolution {\it Chandra} spectroscopy and photometry reveal spectacularly strong X-ray pulsations in the SSS of Nova Her 2021 and remarkably fast outflows with two dominant velocity components at up to 11000~km~s$^{-1}$. Reported pre- and post-outburst optical pulsation periods are significantly different. They could potentially be explained by an expanded envelope and moment of inertia change. Instead, if interpreted in terms of angular momentum loss due to mass ejection, the optical period change implies $2\times 10^{-5}$--$2\times 10^{-4} M_\odot$ was lost. The X-ray pulsation period, with a pulsed fraction of about 60\%\ \added{and a strongly varying pulse profile}, \replaced{is}{appears to be} longer than the reported contemporaneous post-outburst optical pulsation period, \added{although the {\it Chandra} data are also suggestive of the presence of a QPO with period variations similar to the optical and X-ray period differences}. \added{In either case} we speculate that this could be due to different X-ray and optical photospheric depths, rotating non-rigidly relative to the underlying WD.

\begin{acknowledgments}
JJD was supported by NASA contract NAS8-03060 to the {\it Chandra X-ray Center} and thanks the Director, Pat Slane, for continuing advice and support. We also thank the {\it Chandra} Mission Planning group for their work that enabled this DDT observation to be executed. GJML is member of the CIC-CONICET (Argentina) and acknowledges support from grant ANPCYT-PICT 0901/2017. DPKB is supported by a CSIR Emeritus Scientist grant-in-aid
which is being hosted by the Physical Research Laboratory, Ahmedabad. KLP and APB acknowledge funding from the UK Space Agency.
\end{acknowledgments}

\bibliography{v1674her}{}
\bibliographystyle{aasjournal}



\listofchanges

\end{document}